\documentclass[a4paper]{article}

\usepackage[english]{babel}
\usepackage[utf8]{inputenc}
\usepackage{amsmath}
\usepackage{graphicx}
\usepackage[colorinlistoftodos]{todonotes}
\usepackage{subcaption}

\title{The ``Hard Problem'' of Life}

\author{Sara Imari Walker$^{1,2,3,*}$ and Paul C.W. Davies$^1$}

\date{}

\begin{document}
\maketitle
$^1$ Beyond Center for Fundamental Concepts in Science, Arizona State University, Tempe AZ USA\\
$^2$ School of Earth and Space Exploration, Arizona State University, Tempe AZ USA\\
$^3$ Blue Marble Space Institute of Science, Seattle WA USA\\
$^*$ sara.i.walker@asu.edu\\

There are few open problems in science as perplexing as the nature of life and consciousness. At present, we do not have many scientific windows into either. In the case of consciousness, it seems evident that certain aspects will ultimately defy reductionist explanation, the most important being the phenomenon of qualia -- roughly speaking our subjective experience as observers. It is {\it a priori} far from obvious why we should have experiences such as the sensation of the smell of coffee or the blueness of the sky. Subjective experience isn't necessary for the evolution of intelligence (we could for example be zombies in the philosophical sense and appear to function just as well from the {\it outside} with nothing going on {\it inside}). Even if we do succeed in eventually uncovering a complete mechanistic understanding of the wiring and firing of every neuron in the brain, it might tell us nothing about thoughts, feelings and what it is like to experience something. Our phenomenal experiences are the only aspect of consciousness that appears as though it cannot, {\it even in principle}, be reduced to known physical principles. This led Chalmers to identify pinpointing an explanation for our subjective experience as the ``hard problem of consciousness'' \cite{chalmers1995facing}. The corresponding ``easy problems'' (in practice not so easy) are associated with mapping the neural correlates of various experiences. By focusing attention on the problem of subjective experience, Chalmers highlighted the truly inexplicable aspect of consciousness, based on our current understanding. The issue however is by no means confined to philosophy. Chalmers own proposed resolution is to regard subjective consciousness as an irreducible, fundamental property of mind, with its own laws and principles. Progress can be expected to be made by focusing on what would be required for a theory of consciousness to stand alongside our theories for matter, even if it turns out that something fundamentally new is not necessary. 

The same may be true for life. With the case of life, it seems as though we have a better chance of understanding it as a physical phenomenon than we do with consciousness.  It may be the case that new principles and laws will turn out to be unnecessary to explain life, but meanwhile their pursuit may bring new insights to the problem \cite{croninwalker}. Some basic aspects of terrestrial biology, for example, replication, metabolism and compartmentalization, can almost certainly be adequately explained in terms of known principles of physics and chemistry, and so we deem explanations for these features to belong to the ``easy problem'' of life. Research on life's origin for the past century, since the time of Oparin and Haldane and the ``prebiotic soup'' hypothesis, has focused on the easy problem, albeit with limited progress. The more pressing question of course is whether all properties of life can in principle be brought under the ``easy'' category, and accounted for in terms of known physics and chemistry, or whether certain aspects of living matter will require something fundamentally new. This is especially critical in astrobiology, without an understanding of what is meant by ``life'' we can have little hope of solving the problem of its origin or to provide a general-purpose set of criteria for identifying it on other worlds. As a first step in addressing this issue we need to clarify what is meant by the ``hard problem'' of life; that is, to identify which aspects of biology are likely to prove refractory in attempts to reduce them to known physics and chemistry, in the same way that Chalmers identified qualia as central to the hard problem of consciousness. To that end we propose that {\it the hard problem of life is the problem of how `information' can affect the world}. In this essay we motivate both why the problem of information is central to explaining life and why it is hard, that is, why we suspect that a full resolution of the hard problem will not ultimately  be reducible to known physical principles \cite{walkerfundamental}.  

\section{Why this `hard problem'?}

There is an important distinction between the hard problem of life and that of consciousness.  With consciousness it is obvious to each of us that we experience the world -- to read this page of text you are {\it experiencing} a series of mental states, perhaps a voice reading aloud in your head or a series of visual images. The universal aspects of experience are therefore automatically understood to each of us: if intelligent aliens exist and are also conscious observers like us we might expect the objective fact that they experience the world to be similar (even if the experience itself is subjectively different), despite the fact that we can't yet explain what consciousness is or why it arises. By contrast, there is no general agreement on what features of life are universal. Indeed, they could be so abstract that we have yet to identify them \cite{DaviesWalker}. 

As Monod emphasized \cite{monod1974chance}, biological features are a combination of chance and necessity, combining both frozen accidents and law-like evolutionary convergence.  As a result of our anthropocentric vantage point \cite{carter1983anthropic} (thus far observing life only here on Earth), both astrobiology and our assumptions about non-human consciousness tend to be biased by our understanding of terrestrial life. With only one sample of life at our disposal, it is hard to separate which features are merely accidental, or incidental, from the `law-like' features that we expect would be common to {\it all life} in the universe.
 
Discussions about universal features of life typically focus on chemistry. In order to generalize ``life as we know it'' to potential universal signatures of life, we propose to go beyond this emphasis on a single level (chemistry) and recognize that {\it life might not be a level-specific phenomenon} \cite{walker2012evolutionary}. Life on Earth is characterized by hierarchical organization, ranging from the level of cells, to multicellular organisms, to eusocial and linguistic societies \cite{SM94}. A broader concept of life, and perhaps one that is therefore more likely to be universal, could be applied to multiple levels of organization in the biosphere -- from cells to societies -- and might in turn also be able to describe alien life with radically different chemistries. The challenge is to find universal principles that might equally well describe any level of organization in the biosphere (and ones yet to emerge, such as speculated transitions in social and technological systems that humanity is currently witnessing, or may one day soon witness). Much work has been done attempting to unify different levels of organization in biological hierarchies (see {\it e.g., \cite{SM94, campbell1974downward}}), and although we do not yet have a unified theory, many authors have pointed to the concept of information as one that holds promise for uncovering currently hidden universal principles of biology at any scale of complexity ({\it e.g.} \cite{SM94, jablonka1995evolution, FlaErwEll13, smith2008thermodynamics, walker2012evolutionary, farnsworth2013living, DaviesWalker} to name but a few) -- ones that in principle could be extrapolated to life on other worlds. 

Although we do not attempt in this essay to define `biological information' (which is a subject of intense debate in its own right \cite{godfrey2007biological}), we wish to stress that it is not a passive attribute of biological systems, but plays an active role in the execution of biological function (see {\it e.g.} Stotz and Griffiths, Chapter 15 of this volume).  An example from genomics is an experiment performed by the Craig Venter Institute, where the genome from one species was transplanted to another, and `booted up' to convert the host species to the foreign DNA's phenotype -- quite literally re-programming one species into another \cite{lartigue2007genome}. Here it seems clear that it is the {\it information} content of the genome -- the sequence of bits -- and not the chemical nature of DNA as such, which is (at least in part) `calling the shots'. Of course, a hard-nosed reductionist might argue that, in principle, there must exist a purely material narrative of this transformation, cast entirely in terms microstates ({\it e.g.} events at the molecular level). However, one might describe this position as ``promissory reductionism'', because there is no realistic prospect of ever attaining such a complete material narrative, or of its being any use in achieving an understanding of the process even if it was attained. On practical grounds alone, we need to remain open to the possibility that the causal efficacy of information may amount to more than a mere methodological convenience, and might represent a new causal category not captured in a microstate description of the system. What we term ``the hard problem of life'' is the identification of the actual physical mechanism that permits information to gain causal purchase over matter.  This view is not accommodated in our current approaches to physics.   

\section{What is possible under the known laws of physics?} 

Living and conscious systems attract our attention because they are highly remarkable and very special states of matter. In the words of the Harvard chemist George Whitesides 
\begin{quote}
{\it ``How remarkable is life? The answer is: very. Those of us who deal in networks of chemical reactions know of nothing like it? How could a chemical sludge become a rose, even with billions of years to try?''} \cite{Whitesides}
\end{quote}
The emergence of life and mind from non-living chemical systems remains one of the great outstanding problems of science. Whatever the specific (and no doubt convoluted) details of this pathway, we can agree that it represents a transition from the mundane to the extraordinary. 

In our current approaches to physics, where the physical laws are fixed, any explanation we have for why the world is such as it is ultimately boils down to specifying the initial state of universe. Since the time of Newton, our most fundamental theories in physics have been cast in a mathematical framework based on specifying an initial state and a deterministic dynamical law. Under this framework, while physical states are generally time-dependent and contingent, the laws of physics are regarded as timeless, immutable and universal. Although the immutability of the laws of physics is occasionally challenged \cite{wheeler1983recognizing, davies2008goldilocks, smolin2013time, peirce1982writings} it remains the default assumption among the vast majority of scientists. To explain a world as complex as ours -- which includes things like bacteria and at least one technological civilization with {\it knowledge} of things like the laws of gravitation \cite{walkermath} -- requires that the universe have a very special initial state indeed.  The degree of `fine-tuning' necessary to specify this initial state is unsatisfactory, and becomes ever more so the more complex the world becomes \footnote{The alternative explanation that everything interesting has arisen as a result of quantum fluctuations is equally unsatisfactory.}. One should hope for a better explanation than simply resorting to special initial conditions (or for that matter stochastic fluctuations, which in many regards are even less explanatory). Indeed there are serious attempts, such as constructor theory, which aim to remove the dependency on initial conditions from our explanations of the world \cite{DEU3} (see Marletto, Chapter 3 of this volume). Resolving this problem of fine-tuning is essential to explaining life since the most  complex structures we know of -- those least explicable from an arbitrary initial state -- are precisely those of interest in our discussion here, that is living, technological and conscious physical systems ({\it e.g.} things like us). 

An new framework for explanation may come from re-considering the kinds of dynamical trajectories permitted by our current theories. Herein we follow an argument outlined by one of us in \cite{walkermath} (where the focus was why our use of mathematics, as an example of particularly powerful information encoding utilized by living systems, can so effectively describe much of reality). In physics, particularly in statistical mechanics, we base many of our calculations on the assumption of metric transitivity, which asserts that a system's trajectory will eventually explore the entirety of its state space -- thus everything that is physically possible will eventually happen. It should then be trivially true that one could choose an arbitrary ``final state'' ({\it e.g.}, a living organism) and ``explain'' it by evolving the system backwards in time choosing an appropriate state at some 'start' time $t_0$ (fine-tuning the initial state).  In the case of a chaotic system the initial state must be specified to arbitrarily high precision. But this account amounts to no more than saying that the world is as it is because it was as it was, and our current narrative therefore scarcely constitutes an explanation in the true scientific sense.

\begin{figure*}
\centering
  \begin{subfigure}[b]{0.4\textwidth}
        \includegraphics[width=\textwidth]{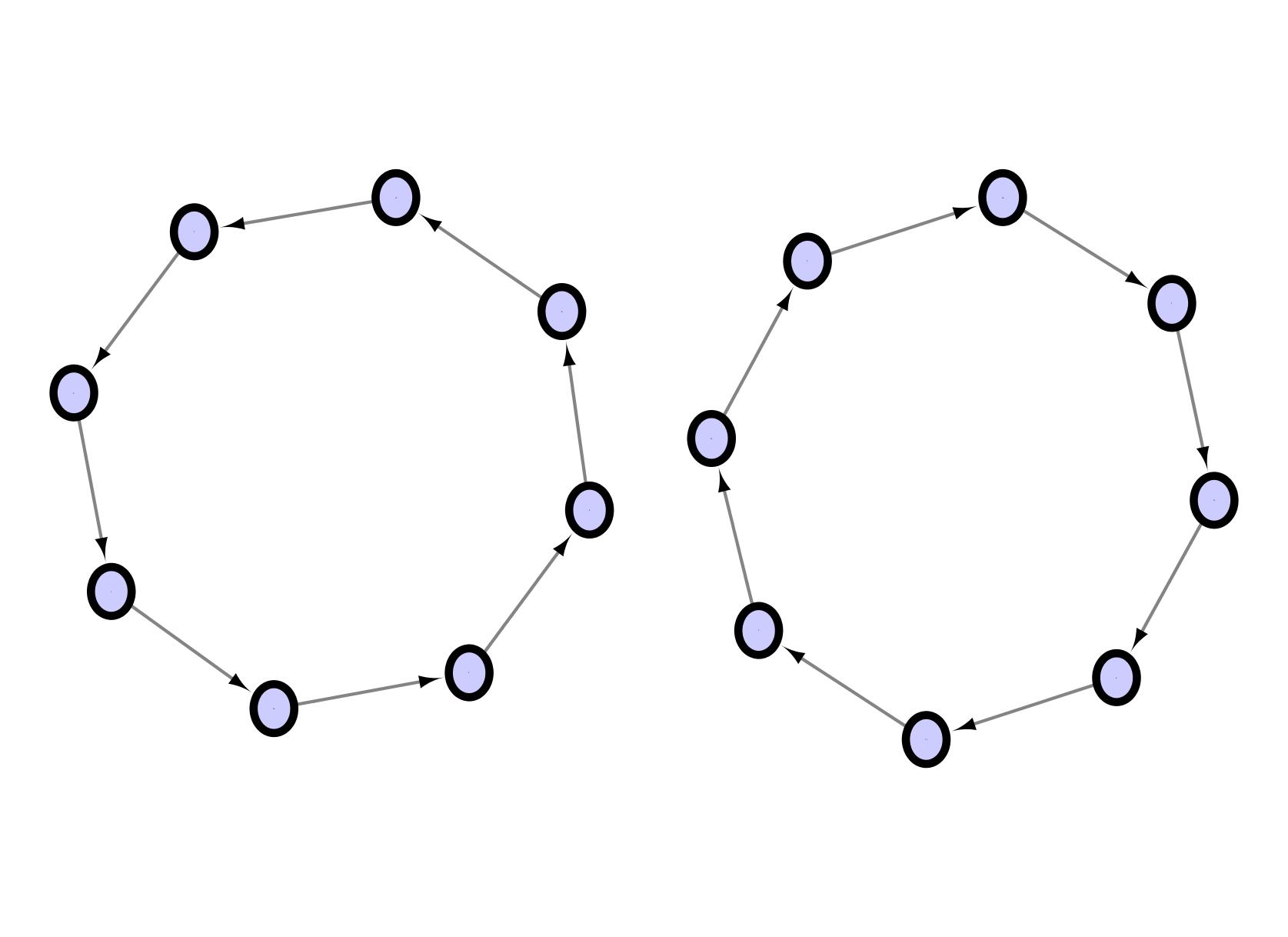}
    \end{subfigure}
    \begin{subfigure}[b]{0.4\textwidth}
        \includegraphics[width=\textwidth]{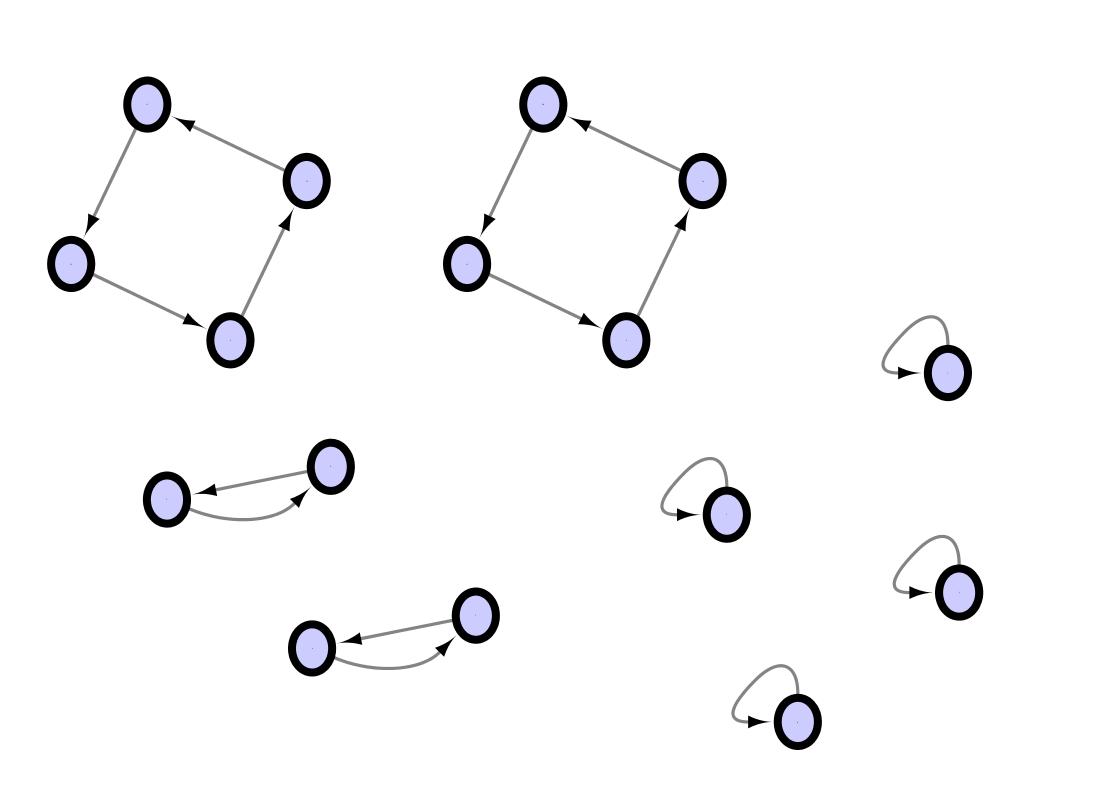}
    \end{subfigure}
\caption{Reversible causal graphs demonstrate that `you can't always get there from here' as the state space is composed of many disconnected regions.} \label{fig:ca_net}
\end{figure*}

We are left in a bit of a conundrum with respect to the problem of specifying the initial conditions necessary to explain our world. A key point is that if we require specialness in our initial state (such that we observe the current state of the world and not any other state) metric transitivity cannot hold true, as it blurs any dependency on initial conditions -- that is, it makes little sense for us to single out any particular state as special by calling it the 'initial' state. If we instead relax the assumption of metric transitivity (which seems more realistic for many real world physical systems -- including life), then our phase space will consist of isolated pocket regions and it is not necessarily possible to get to any other physically possible state (see {\it e.g.} Fig. \ref{fig:ca_net} for a cellular automata example). Thus the initial state must be tuned to be in the region of phase space in which we find ourselves, and there are regions of the configuration space our physical universe would be excluded from accessing, even if those states may be equally consistent and permissible under the microscopic laws of physics (starting from a different initial state). Thus according to the standard picture, we require special initial conditions to explain the complexity of the world, but also have a sense that we should not be on a particularly special trajectory to get here (or anywhere else) as it would be a sign of fine--tuning of the initial conditions. Stated most simply, a potential problem with the way we currently formulate physics is that you can't necessarily get everywhere from anywhere (see Walker \cite{walkermath} for discussion).

A real living system is neither deterministic nor closed, so an attempt to attribute life and mind to special initial conditions would necessarily involve fixing the entire cosmological initial state to arbitrarily high precision, even supposing it were classical. If instead one were to adopt a quantum cosmological view, then the said pathway from the mundane to the extraordinary could of course be accommodated within the infinite number of branches of the wave function, but again this is scarcely a scientific explanation for it merely says that anything that can happen, however extraordinary, will happen somewhere within the limitless array of histories enfolded in the wave function.

Leaving aside appeal to special initial conditions, or exceedingly unusual branches of cosmological wave functions, one may still ask whether pathways from the mundane to the extraordinary are problematic within the framework of known physics. Here we wish to point out a less appreciated fact with respect to the problem of fine-tuning and explaining the complexity of our world. Just because every intermediate state on a pathway to a novel state is physically possible does not mean that an arbitrary succession of states is {\it also possible}. If we envisage the route from mundane chemistry to life, and onward to mind, as a trajectory in some enormous state space, then not every trajectory is consistent with the known laws of physics. In fact, it may well be that almost all trajectories are inconsistent with the known laws of physics (this could be true even if individual steps taken along the way are compatible with known laws).

To justify this claim we explore a toy model inspired by cellular automata (CA), which are often used as computational models for exploring aspects of living and complex systems. However, we note that our arguments, as presented here, are by no means exclusive to CA and could apply to any discrete dynamical system with fixed rules for its time evolution. We note that it is not necessarily the case that the physical laws governing our universe are completely deterministic (for example, under collapse interpretations of quantum theory) and also that reality is not necessarily discrete. However, by demonstrating a proof-of-principle for the more conservative case of discrete deterministic systems we expect that at least some aspects will be sufficiently general to apply to physical laws as they might describe the real universe under relaxed assumptions from those presented herein. 

CA are examples of discrete dynamical systems that consist of a regular grid of cells, each of which can be in a finite number of states -- in particular, we focus on systems with cells that can be in one of two possible states '0' or '1'. For simplicity, let's also assume our universe is one-dimensional with a spatial size of $w$ cells. The configuration space of the system then contains $2^w$ possible states. If we restrict ourselves to deterministic systems, saying nothing yet about the laws that operate on them, each state may appear exactly once on any given trajectory, prior to settling into an attractor (otherwise the system would not be deterministic). Under this constraint, the total number of deterministic trajectories of length $r \leq 2^w$, $n_t(r)$, is just the number of possible permutations on a subset $r$ chosen from $2^w$ elements:
\begin{eqnarray}
n_t (r) = \frac{2^w!}{(2^w - r)!}
\end{eqnarray}
which quantifies the number of ways to obtain an {\it ordered} subset of $r$ elements from a set of $2^w$ elements. The total number of unique possible trajectories is just the sum over all possible trajectory lengths $r$:
\begin{eqnarray}
N = \sum_{r=1}^{2^w} \frac{2^w!}{(2^w - r)!} = e \Gamma (1 + 2^w, 1) - 1  \label{eq:N}
\end{eqnarray}
where $\Gamma(x,a)$ is the incomplete gamma function. The above includes enumeration over all non-repeating trajectories of length $2^w$ and trajectories of shorter length that settle to an attractor at a time $r < 2^w$. Here, {\it $N$ should be interpreted as the number of total possible deterministic trajectories through a configuration space, where states in the space each contain $w$ bits of information}. So far, our considerations are independent of any assumptions about the laws that determine the dynamical trajectories that are physically realized.  We can now consider the number of possible trajectories for a given class of {\it fixed deterministic laws}. 

A natural way to define a `class' of laws in CA is by their locality. For example, elementary cellular automata (ECA) are some of the simplest one-dimensional CA studied, and are defined by a nearest-neighbor interaction neighborhood for each cell, where cell states are defined on the two bit alphabet $\{0,1\}$. Nearest-neighbor interactions define a neighborhood size $L=3$ (such that a cell is updated by the fixed rule, according to its own state and two nearest neighbors), which we herein define as the locality of an ECA rule. For ECA there are $R = 2^{2^L} =256$ possible fixed `laws' (ECA rules) (see \cite{nks}). We can therefore set an upper bound for the number of trajectories contained within {\it any} given rule set, defined by its neighborhood $L$ as:
\begin{eqnarray}
f_L \leq 2^{2^L} \times 2^w \label{eq:n_L}
\end{eqnarray}
where $f_L$ is the total number of possible realized trajectories that {\it any} class of {\it fixed}, deterministic laws operating with a locality $L$ could realize, starting from a set of $2^w$ possible initial states (that is, any initial state with $w$ bits of information)\footnote{This is an upper bound as it assumes each trajectory in the set is unique, but as it happens it is possible, for example, that the application of two different ECA rules in the set of all ECA could yield the same trajectory, so this constitutes an absolute upper bound.}. Eqs. \ref{eq:N} and \ref{eq:n_L} are not particularly illuminating taken alone, instead we can consider the ratio of the upper bound on the number of trajectories possible under a given set of laws to the total number of deterministic trajectories:
\begin{eqnarray} \label{eqn:frac}
\frac{f_L}{N} = \frac{2^{2^L} \times 2^w}{e \Gamma (1 + 2^w, 1) - 1}~.
\end{eqnarray} 
Taking the limit as the system size tends towards infinity, that is as $w \rightarrow \infty$ yields:
\begin{eqnarray}
\lim_{w \to\infty} \left[ \frac{2^{2^L} \times 2^w}{e \Gamma (1 + 2^w, 1) - 1} \right] = 0~.
\end{eqnarray} 
Note that this result is {\it independent} of $L$. {\it For any class of fixed dynamical laws one chooses (any degree of locality), the fraction of possible physically realized trajectories rapidly approaches zero as the number of bits in states of the world increases} \footnote{ This gets worse if states contain more information, that is if the alphabet size $m > 2$.}. Thus, the set of all physical realizations of universes evolved according to local, fixed deterministic laws are very impoverished compared to what could potentially be permissible over all possible deterministic trajectories (and worse so if one considers adding stochastic or non-deterministic trajectories in the summation in Eq. \ref{eq:N}). Only an infinitesimal fraction of paths are even realizable under {\it any} set of laws, let alone by a particular law drawn from any set. 

If we impose time reversal symmetry on the CA update rules, by analogy with the laws of physics, there is an additional restriction:  only $8$ of the $256$ ECA rules are time-reversal invariant.  For these laws, there is no single trajectory that includes all possible states (see Fig. \ref{fig:ca_net}). Thus, we encounter the problem that 'you can't get there from here' and even if you are in the right regime of configuration space, there is only one path (ordering of states) to follow. 

Of course, explanations referring to real biological systems differ from a CA in several respects, not least of which is that one deals with macrostates rather than microstates. However, the conclusion is unchanged if one considers the dynamics of macrostates rather than microstates: the trajectories among all possible macrostates will also be diminished relative to the total number of trajectories (this is because the information in macrostates is less than in the microstates, {\it e.g.} will be $< 2^w$ macrostates for our toy example) \footnote{This holds even if one considers that the number of possible partitions of our state space for $2^w$ possible states is given by the Bell number $B_n$, where $n = 2^w$ which approaches $\infty$ more slowly than the denominator in Eq. \ref{eqn:frac}}. This toy model cautions us that in seeking to explain a complex world which is ordered in a particular way ({\it e.g.}, contains living organisms and conscious observers), based on fixed laws that govern microstate evolution, we may well need to fine-tune not only the initial state, but also the laws themselves in order to specify the particular ordering of states observed (constraining the universe to a unique past and future). Expressed more succinctly, if one insists on attributing the pathway from mundane chemistry to life as the outcome of fixed dynamical laws, then (our analysis suggests) those laws must be selected with extraordinary care and precision, which is tantamount to intelligent design: it states that ``life'' is `written into' the laws of physics {\it ab initio}. There is no evidence at all that the actual known laws of physics possess this almost miraculous property.

The way to escape from this conundrum -- that `you can't get anywhere from here' -- is clear: we must abandon the notion of fixed laws when it comes to living and conscious systems.

\section{Life ... from new physics?}

Allowing both the states {\it and the laws} to evolve in time is one possibility for alleviating the problems associated with fine-tuning of initial states and dynamical laws, as discussed in the previous section. But, this cannot be done in an ad hoc way and still be expected to be consistent with our known laws of physics. A trivial solution would be to assume that the laws are time-dependent and governed by meta-laws, but then one must explain both where the laws and meta-laws come from, and whether there are meta-meta-laws that govern the meta-laws, and meta-meta-meta-laws {\it ad infinitum}. This is therefore no better an explanation than our current framework as it just pushes the problem off to be one of meta-laws rather than laws. As it stands right now, there exists no compelling evidence that any of our fundamental laws depend on time (although claims are occasionally made to the contrary \cite{webb2011}.

\begin{figure*}
\centering
\includegraphics[width=0.75\textwidth]{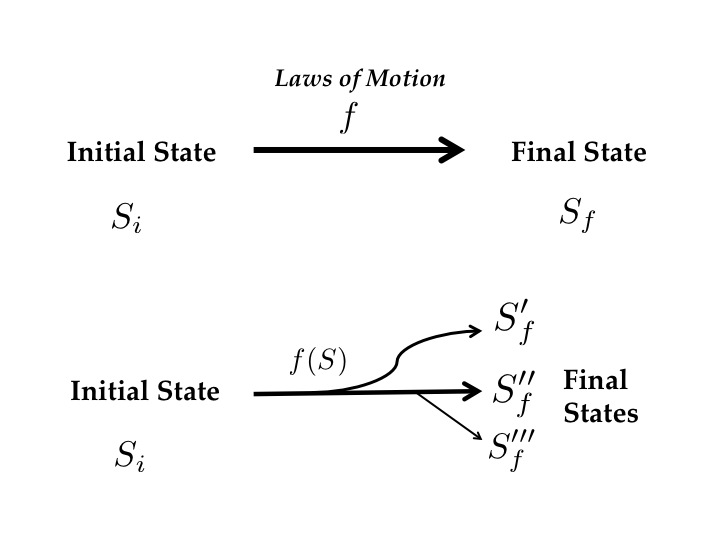}
\caption{Contrast between the fixed dynamical laws of physics (top) and the apparent historical, state-dependent trajectories characteristic of living systems (bottom).} \label{fig:paths}
\end{figure*}

A better idea is to assume that there exist conditions under which the dynamical rules are a function of the states (and therefore are in some sense an emergent property) \cite{Adams2016, pavlic2014self}. Indeed, this seems consistent with what we know of life, where the manner in which biological systems evolve through time is clearly a function of their current state \cite{goldenfeld2010life}. Biological systems appear to be incredibly path-dependent in their dynamical trajectories, as exemplified by the process of biological evolution. Starting from the same initial state (roughly speaking a genome), biological systems trace out an enormous array of alternative trajectories through evolutionary adaptation and selection (see Fig. \ref{fig:paths}). It is difficult, if not impossible, to write an equation of motion with a fixed rule for such historical processes due to their state-dependent nature. State-dependent laws are a hallmark of self-referential systems such as life and mind \cite{goldenfeld2010life, WalkerDavies2013, hofstadter1980godel}.  It is a contrast with our current views of immutable laws in physics keenly appreciated by Darwin, as eloquently expressed in the closing passage of his book {\it On the Origin of Species} \cite{darwin1991origin}:
\begin{quote}
{\it ``whilst this planet has gone cycling on according to the fixed law of gravity, from so simple a beginning endless forms most beautiful and most wonderful have been, and are being, evolved.''}
\end{quote}

It seems that one resolution to the above conundrum, consistent with what we know of biology, is therefore to introduce {\it state-dependent dynamical laws}. However, there is a trivial sense in which state-dependent dynamical laws might alleviate the fine-turning problem: that is, by assuming that each state uniquely specifies the next state. One could then construct a state-dependent algorithm unique to each possible trajectory, such that there is exactly one algorithm that would ``explain'' each path taken to reach any arbitrary final state. However, this is inconsistent with what we know of physics: such laws would not be algorithmically compressible and therefore it would be impossible to write such a succinct equation as $F = ma$ to describe {\it anything} about a universe governed by these kinds of laws. 

How then can we reconcile the biological narrative with our current understanding of physics? When we externally describe a system, we articulate the counterfactual possibilities and assign a quantity of ``information'' to them. However, in order for these counterfactual possibilities to be physically realized, the information specifying them must be contained within the system, and contribute to specifying which dynamical path through state space is taken (consistent with the underlying physical laws). This should be a local property, yielding an effective description that is state-dependent. The challenge is that we do not have a physical theory for information that might explain how information could ``call the shots''. There are some indications for a potentially deep connection between information theory (which is not cast as a physical theory and instead quantifies the efficacy of communication through noisy channels), and thermodynamics, which is a branch of physics\footnote{Stating that information theory is not a physical theory is not the same as to say that information is not physical -- a key insight of information theory is that information is a measurable physical quantity. ``Information is physical!'' in the words of Rolf Landauer\cite{landauer1996physical}.} due to the mathematical relationship between Shannon and Boltzmann entropies. Substantial work over the last decade has attempted to make this connection explicit, we point the reader to \cite{Parrondo2015, lutz2015information} for recent reviews. Schr\"odinger was aware of this link in his deliberations on biology, and famously coined the term ``negentropy'' to describe life's ability to seemingly violate the $2^{nd}$ law of thermodynamics\footnote{``Schr\"odinger's paradox'' with regard to life's ability to generate `negative entropy' is quickly resolved if one considers that living systems are open to an environment}. Yet he felt that something was missing, and that thermodynamic considerations alone are insufficient to explain life \cite{schrodinger2004mind}:

\begin{quote}
{\it ``$\ldots$ living matter, while not eluding the "laws of physics" as established up to date, is likely to involve "other laws of physics" hitherto unknown \ldots''}
\end{quote}

We suggest one approach to get at these ``other laws'' is to focus on the connection between the concept of ``information'' and the equally ill-defined concept of ``causation'' \cite{walkerinformational, kim2015new, DaviesWalker2016}. Both concepts are implicated in the failure of our current physical theories to account for complex states of the world without resorting to very special initial conditions.  In particular, we posit that the manner in which biological systems implement state-dependent dynamics is by utilizing information encoded {\it locally} in the current state of the system, that is, by attributing causal efficacy to information. It is widely recognized that coarse-graining (which would define the relevant `informational' degrees of freedom) plays a foundational role in how biological systems are structured \cite{FlaErwEll13}, by defining the biologically relevant macrovariables (see {\it e.g.} Chapters by Flack, Dedeo and by Wolpert {\it et al.} in this volume). However, it is not clear how those macrostates arise, if they are objective or subjective \cite{shalizi2003macrostate}, or whether they are in fact a fundamental aspect of biological organization -- {\it intrinsic to the dynamics} (i.e. such that macrostates are causal) rather than merely a useful phenomenological descriptor.  A framework in which coarse-grained information-encoding macrostates are causal holds promise for resolving many of the problems discussed herein. This is the key aspect of the hard problem of life.

\section{Conclusions}

There are many difficult open problems in understanding the origin of life -- such as the `tar paradox' \cite{benner2014paradoxes} and the fact that prebiotic chemistry is just plain hard to do. These problems differ qualitatively from the `hard problem of life' as identified here. Most open problems associated with life's origin such as these, while challenging right now, will likely ultimately reduce to known principles of physics and chemistry and therefore constitute by our definition ``easy problems''. Here we have attempted to identify a core feature of life that won't similarly be solved based on current paradigms -- namely, that life seems distinct from other physical systems in how information affects the world (that is, that macrostates are causal). We have focused on the problem of explaining the pathway from non-living chemical systems to life and mind to explicate this problem and have attempted to motivate why new principles and potentially even physical laws are necessary. One might regard this as too a radical step, however it holds potential for resolving deep issues associated with what life is and why it exists. Previous revolutions in our understanding of physical reality, such as general relativity and quantum mechanics, dramatically reshaped our understanding of the world and our place in it. To quote Einstein, `One can best feel in dealing with living things how primitive physics still is.' ( A. Einstein, letter to L. Szilard quoted in \cite{prigogine1997end}). Given how much more intractable life seems, we should not immediately jump to expecting anything less of a physical theory that might encompass it. If we are so lucky as to stumble on new fundamental understanding of life, it could be such a radical departure from what we know now that it might be left to the next generation of physicists to reconcile the unification of life with other domains of physics, as we are now struggling to accomplish with unifying general relativity and quantum theory a century after those theories were first developed. 

\section*{Acknowledgements} This work was made possible through support of a grant from Templeton World Charity Foundation. The opinions expressed in this publication are those of the author(s) and do not necessarily reflect the views of Templeton World Charity Foundation.

\bibliographystyle{abbrv}
\bibliography{Ref_master}

\end{document}